\documentclass{iau} 
\usepackage{graphicx}

\title[Are some meteoroids rubble piles?] %% give here short title %%
{Are some meteoroids rubble piles?}

\author[J. Borovi\v{c}ka]   %% give here short author list %%
{Ji\v{r}\'{\i} Borovi\v{c}ka$^1$}

\affiliation{$^1$Astronomical Institute of the Czech Academy of Sciences, Fri\v{c}ova 298, CZ-25165 Ond\v{r}ejov, 
Czech Republic \\email: {\tt jiri.borovicka@asu.cas.cz}}

\pubyear{2015}
\volume{318}  %% insert here IAU Symposium No.
\jname{Asteroids: New Observations, New Models}
\editors{S. Chesley, A. Morbidelli, \& R. Jedicke, eds.}

\begin{document}

\maketitle

\begin{abstract}
The possibility that some meteoroids in the size range 1 -- 20 meters are rubble piles i.e.\ 
assembles of boulders of various sizes held together only by small van der Waals forces,
is investigated. Such meteoroids are expected to start disrupting into individual pieces during
the atmospheric entry at very low dynamic pressures of $\sim 25$ Pa, even before the onset
of ablation. The heterogeneous bodies as Almahata Sitta (asteroid 2008 TC$_3$) and Bene\v{s}ov are
primary candidates for rubble piles. Nevertheless, by analyzing the deceleration, wake, and light curve of the Bene\v{s}ov bolide, 
we found that the meteoroid disruption started only at a height of 70 km under dynamic pressure of 50 kPa.
No evidence for a very early fragmentation was found also for the Chelyabinsk event.

\end{abstract}

\firstsection % if your document starts with a section,
\section{Introduction}

It is now commonly accepted that most asteroids in the size range 200 m -- 10 km are rubble piles, i.e.\ 
assembles of boulders of various sizes held together only by mutual gravity. The main evidence for this 
is the existence of the surface disruption spin limit (\cite[Pravec \& Harris 2000]{Pravec}, \cite[S\'anchez
\& Scheeres 2014]{SS}). The rotation periods of asteroids in this size range are in almost all cases
longer than 2.3 hours, corresponding to the limit at which the centrifugal force at the surface equals 
to the gravitational force. Rubble piles are products of asteroid collisions leading to disruption of bodies and
re-assembly of fragments.

The  rotation periods of
asteroids smaller than 200 m are often shorter
than the surface disruption spin limit, sometimes shorter than one minute. These bodies were therefore 
considered to be mostly monolithic with significant strength. Nevertheless, \cite[S\'anchez
\& Scheeres (2014)]{SS}  considered small van der Waals forces between the grains inside 
rubble piles. They found that the strength of rubble piles may be about 25 Pa and that this low strength is sufficient to hold
together small asteroids with rotational periods of the order of minutes. In particular they argued that
the asteroid 2008 TC$_3$ may have been a rubble pile despite of its rotational period 
of 99 seconds. 

Asteroid 2008 TC$_3$ was discovered 19 hours before it impacted the Earth on October 7, 2008 
(\cite[Jenniskens et al. 2009]{Jenn}). Photometric observations before the impact revealed that 
the asteroid was an elongated body in excited rotational state with period of rotation 99.2 s and 
period of precession 97.0 s (\cite[Scheirich et al. 2010]{Scheirich}). By combining various data, the most 
probable dimensions were estimated to be $6.6\times3.6\times2.4$ m, mass 40,000 kg, bulk density 1800 kg m$^{-3}$, 
and porosity $\sim 50\%$ (\cite[Borovi\v{c}ka et al. 2015]{AIV}).
The impact occurred in Sudan and numerous small meteorites ($< 0.4$ kg) were found in the desert
(\cite[Jenniskens et al. 2009]{Jenn}). Surprisingly, the meteorites were of various
mineralogical types (\cite[Bischoff et al. 2010]{Bischoff}, \cite[Shaddad et al. 2010]{Shaddad}).
The body was therefore clearly heterogeneous and seems to be good candidate for
a rubble pile. The data on the behavior of the body
during the atmospheric entry are, unfortunately, scarce. There was a major flare at the height of 37~km
and probably other flares at 53, 45, and 32 km (\cite[Jenniskens et al. 2009]{Jenn}, \cite[Borovi\v{c}ka \& 
Charv\'{a}t 2009]{Meteosat}). 

Fireball flares are evidences of meteoroid fragmentation. Atmospheric fragmentation of meteoroids 
is a common process (e.g. \cite[Ceplecha et al. 1993]{Cepl93}). It occurs when
the dynamic pressure, $p=\rho v^2$ ($\rho$ is atmospheric density and $v$ is meteoroid
velocity) exceeds meteoroid strength. While the tensile strength of monolithic rocks 
(meteorites) exceeds 30 MPa, the strengths of meteoroids inferred from their atmospheric
fragmentations was found to be in the range 0.1 -- 10 MPa (\cite[Popova et al. 2011]{Popova}). 
The lowered strength is likely caused by internal fractures.  In this respect Almahata Sitta was not exceptional. The
flares occurred at pressures 0.3 -- 1.3 MPa. Such strength is
much higher than 25 Pa expected for rubble piles.

In this paper I explore the possibility that the observed fragmentations are in fact
only the secondary break-ups of the building boulders of rubble piles. At least in some cases
the first break-up may occurs at pressures of $\sim25$ Pa. 
The question is if we can find evidences for the initial high altitude fragmentation in the bolide data.

\begin{figure}
\vspace*{-0.4 cm}
\begin{center}
\includegraphics[width=3.2in]{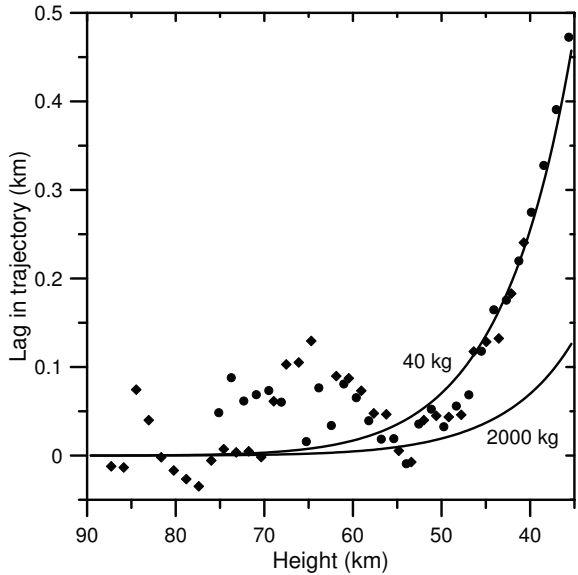} 
\vspace*{-0.2 cm}
\caption{Deceleration in the middle part of the Bene\v{s}ov bolide. The lag in trajectory is plotted
as a function of height. Dots and diamonds are measurements from two spectral cameras. The curves show the
computed lag for meteoroids of masses 2000 kg and 40 kg, respectively 
(assuming $\Gamma=0.6$, $A=1.21$, $\rho_d=3000$ kg m$^{-3}$).}
\label{decel}
\end{center}
\end{figure}
\section{Bene\v{s}ov}

Since there are no detailed data on the Almahata Sitta bolide, I will inspect another good
candidate for rubble pile -- the Bene\v{s}ov meteoroid. The Bene\v{s}ov meteoroid entered the
atmosphere over the Czech Republic on May 7, 1991 (\cite[Spurn\'y 1994]{Ben94}). The bolide was well observed by
three all-sky cameras and two high resolution photographic spectrographs (\cite[Borovi\v{c}ka
and Spurn\'y 1996]{Ben96}). Four small meteorites were recovered in 2011 -- 2012. The meteorites
were of different mineralogical types (H and LL chondrites with achondritic clast), 
similarly to Almahata Sitta (\cite[Spurn\'y et al. 2014]{Ben14}). The initial mass of the
meteoroid, derived primarily from the amount of radiated energy, was 2000 -- 4000 kg 
(\cite[Borovi\v{c}ka et al. 1998]{Ben98}, \cite[Ceplecha \& ReVelle 2005]{CepReV}). 
The diameter was therefore larger than one meter.
The initial velocity was 21.3 km s$^{-1}$ and the trajectory was almost vertical. Obvious
fragmentations occurred at heights 38 km and 25 km, under dynamic pressures of 2.5 MPa
and 9 MPa, respectively. Nevertheless, the deceleration at heights around 45 km was so strong
that the body must have been fragmented already there.

\begin{figure}
\vspace*{-0.4 cm}
\begin{center}
\vspace*{1mm}
\includegraphics[width=4.8in]{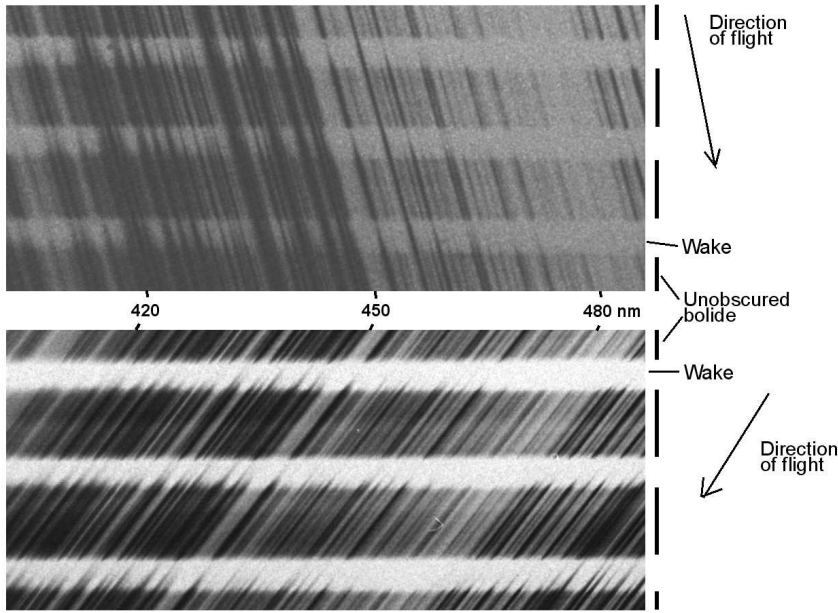} 
\vspace*{-0.6 cm}
\caption{Parts of two spectral plates showing the Bene\v{s}ov spectrum at heights around 70 km (top) and 45 km (bottom). 
Wake is seen between segments of unobscured bolide. Approximate wavelengths in nm are indicated.}
\label{spectrum}
\end{center}
\end{figure}

Figure~\ref{decel} demonstrates the observed deceleration. The lag in trajectory is plotted as
a function of height. The lag is the difference between the actual position of the fireball
at a given time and the position expected for constant velocity.
The lag is zero as long as there is no deceleration. Naturally, at a given height, 
deceleration will be larger for a smaller body (provided that shape and density is the same). 
Surprisingly, the observed lag does not follow the curve for a mass
of 2000 kg. Instead, the mass corresponding to the observed lag is only 40 kg.
The discrepancy was noted already by \cite[Borovi\v{c}ka et al. (1998)]{Ben98} but
the mass was then computed for $\Gamma A=1.0$ ($\Gamma$ is the drag coefficient and 
$A$ is the shape coefficient). Here we use a more realistic value $\Gamma A = 0.7$.
The density of the meteoroid is assumed to be $\rho_d=3000$ kg m$^{-3}$. Lowering
the density to 400 kg m$^{-3}$ would explain the observed deceleration, however, such a low
density is unrealistic considering the types of the meteorites. It is much more likely
that the meteoroid was already disrupted into large number of fragments at a height of
50 km. The mass of the largest fragment was about 40 kg.

The question is where the disruption occurred. The dynamic pressure of 25 Pa was reached
at the height of 113 km, while the bolide started to be visible only at the height of 91 km.
In principle it can be possible that the initial disruption occur earlier than
the meteoroid surface reach the temperature needed for ablation and radiation. In that case we would not see
a direct evidence for fragmentation height in the bolide data. Nevertheless, it can be
expected that fragments of various masses are formed in the disruption. Mass segregation
then occurs since smaller fragments decelerate more. At lower heights, the fireball 
will not be a point-like object but will be elongated with wake formed by smaller fragments. 
The length of the wake will depend on the mass distribution of fragments and on the height
of fragmentation. The earlier the fragmentation occurred, the longer will be the wake.

To investigate Bene\v{s}ov wake, we can use the high resolution spectral photographs. They were taken
with lenses of focal length of 360 mm. The spatial resolution at bolide range 100 km is of the order of several meters.
Each camera was equipped by a transmission diffraction grating in front of the lens. As it is usual for bolides, 
the spectrum consisted primarily of atomic lines of metals evaporated from the meteoroid, in particular Fe, Mg, Na,
Cr, Mn, and Ca. For our purposes it is important that the cameras were periodically closed by a rotating shutter.
The frequency was 15 Hz and the open-to-close ratio was approximately 2:1. If the bolide were point-like, no signal would
be visible between the shutter breaks. This was the case at the beginning, at heights above 84 km. 

At lower heights, strong wake developed. Figure~\ref{spectrum} shows parts of the spectra at heights 70 km and 45 km. 
The wake at 70 km was so long that it filled the whole gaps between the shutter breaks. However, this was the case
only for certain spectral lines, in particular low excitation lines of Fe and Mg with low transition probability. These lines
are known to be strong in meteor wakes (\cite[Halliday 1968]{Halliday}). That kind of wake is, however, not produced
by fragments but by cooling rarified gas behind the meteoroid. The situation changed at lower heights. At 45 km, the
wake was shorter but its spectrum was more similar to the spectrum of the bolide head (Fig.~\ref{spectrum}). 	
We suppose that the wake was produced here mainly be small fragments lagging behind the large fragments forming
the head.

The length of the wake was about 250 meters at the height of 50 km. 
Such length can be explained by fragments of masses 40 -- 0.1 kg separated at 65 km. 
If separated at 113 km, the mass range must have been narrower, 40 -- 0.5 kg.
While the larger mass range is more likely, the difference between these two scenarios is not substantial.
The reason is that deceleration at heights above 80 km is negligible (even for gram-sized fragments). The length of
the wake is therefore, unfortunately, not very sensitive to the actual height of disruption.

\begin{figure}
\vspace*{-0.4 cm}
\begin{center}
\includegraphics[width=3.6in]{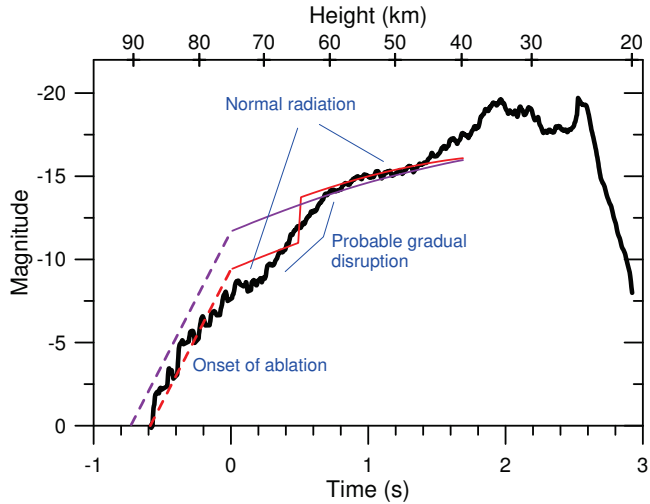} 
\vspace*{-0.2 cm}
\caption{Measured light curve of the Bene\v{s}ov bolide (thick black line) and two models of the middle of the bolide. The red
curve is a model of sudden disruption at the height of 65 km. The violet curve is for disruption before the start of ablation.
The onset of ablation (dashed parts) is shown only schematically.}
\label{light}
\end{center}
\end{figure}

We will now look more closely at the light curve of the Bene\v{s}ov bolide. Figure~\ref{light} shows the light curve as
measured by \cite[Borovi\v{c}ka \& Spurn\'y (1996)]{Ben96}. Shown are also two models of the ascending part of the
light curve. There were no flares in this part, which would point out to fragmentation events. However, the slope changed
several times. The steep slope at the beginning (time $<0$ s) can be ascribed to the onset of ablation. This part was not
modeled in detail. Our model (described in  \cite[Borovi\v{c}ka et al. 2013a]{Kosice}) 
assumes that the ablation is in full progress. In that case the slope of the light curve in the middle part of the bolide is almost
constant. Indeed, the observed slope corresponds to the modeled slope during two intervals: 0 -- 0.2 s (heights 75 -- 70 km) and 0.7 -- 1.4 s
(60 -- 45 km). In between, the slope was steeper. The
model of instantaneous disruption at the height of 65 km shows a step on the light curve with the
increase of brightness by more than two magnitudes at 65 km. The step is due to increased cross section of the meteoritic material
after disruption. In reality, the disruption was more gradual and occurred within 0.5 s at heights 70 -- 60 km. The second
model, which assumed that the meteoroid had been disrupted already before the start of ablation, predicts too bright
bolide at heights above 60 km and cannot explain the observed change of slope.

We therefore conclude that the disruption of Bene\v{s}ov meteoroid started at the height of 70 km. The dynamic pressure at that
time was 50 kPa, i.e.\ three orders of magnitude higher than the strength of rubble piles. Although not a rubble pile, the bulk
strength of Bene\v{s}ov was lower than of other meteorite dropping meteoroids (\cite[Popova et al. 2011]{Popova}). The
initial disruption was severe -- the largest fragments were of only 1 -- 2\% of the original mass ($\sim 25$\% in terms of size).
But the low strength is in accordance with the heterogeneous nature of the recovered meteorites 
(\cite[Spurn\'y et al. 2014]{Ben14}).

Note that the similarly massive  \v Sumava meteoroid fragmented at similar pressures, namely
25 -- 140 kPa  (\cite[Borovi\v{c}ka \& Spurn\'y 1996]{Ben96}).
The behavior was, however, completely different in that case. 
The body was completely destroyed at height 59 km after several disruptions 
accompanied by large amplitude flares. \v Sumava was likely a cometary body with extremely high microporosity
and low density ($\sim$ 100 kg m$^{-3}$) and easily disintegrated into dust.

\section{Chelyabinsk}

\begin{figure}
\vspace*{-0.4 cm}
\begin{center}
\includegraphics[width=3.4in]{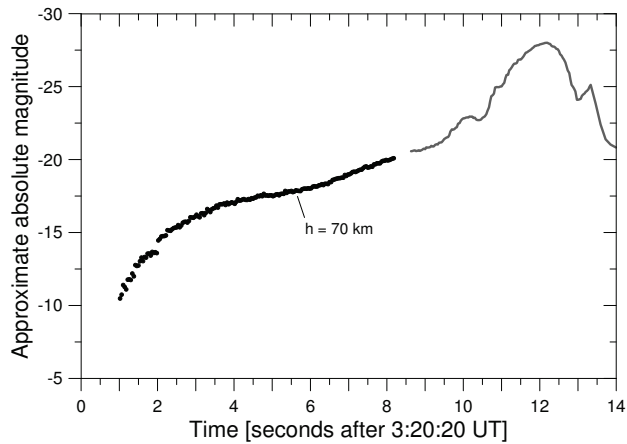} 
\vspace*{-0.2 cm}
\caption{Combined light curve of the Chelyabinsk bolide. The part of the right (bright phase) was taken from
\protect\cite[Brown et al. (2013)]{Nature2}. The part of the left was obtained (on relative scale) by measuring the bolide signal on 
Beloretsk video (video no.\ 14 in \protect\cite[Borovi\v cka et al., 2013b]{Nature1}), taking into account bolide range, 
atmospheric extinction, and image saturation.}
\label{chelc}
\end{center}
\end{figure}

We can also briefly look at the Chelyabinsk event of 15 February 2013 -- the largest well observed impact
(\cite[Brown et al. 2013]{Nature2}, \cite[Popova et al. 2013]{Science}). The impactor size was $\sim 19$ meters (mass $10^7$ kg), the
entry speed was
19 km s$^{-1}$, and trajectory slope 18$^\circ$. The first obvious fragmentation occurred at a height of 45 km under
dynamic pressure of 0.5 MPa. Catastrophic disruption occurred at 1 -- 5 MPa 
(\cite[Borovi\v{c}ka et al. 2013b]{Nature1}, \cite[Popova et al. 2013]{Science}). Deceleration was negligible until the disruption, yielding
a lower limit of the mass before the disruption of $10^6$ kg (\cite[Borovi\v{c}ka et al. 2013b]{Nature1}). From that we cannot
say if some high altitude fragmentation occurred or not. Wake was presented already at height 85 km but we do not have spectra
to judge its nature. A lot of dust was deposited in the atmosphere. The massive dust trail started already at height 70 km
(dynamic pressure of 25 kPa).
We inspected the light curve 
(Fig.~\ref{chelc}) 
and no flare and no change of slope was found at heights around 70 km. There is therefore no evidence
for an early fragmentation. The dust was likely lost from the surface of the body.

\section{Conclusions}

There is no evidence so far of a meteoroid in the 1 -- 20 meter size range being a rubble pile.
Most meteoroids are fractured rocks with strengths of 0.1 -- 10 MPa.
Even the heterogeneous bodies (Bene\v{s}ov, 2008 TC$_3$) had strength $> 10$ kPa. So, there must be a mechanism stronger than
van der Waals forces to hold the reaccumulated bodies together.
We, however, note that early fragmentation during the atmospheric entry may not be easy to recognize in all cases.

\acknowledgements 
This work was supported by project P209/11/1382 from the Czech Science Foundation (GA\v{C}R).
The institutional project was RVO:67985815.


\begin{thebibliography}{}

\bibitem[Bischoff et al. (2010)]{Bischoff}
{Bischoff, A., Horstmann, M., Pack, A., Laubenstein, M., \& Haberer S.} 2010.
\textit{Meteorit. Plan. Sci.}, 47, 1638

\bibitem[Borovi\v{c}ka \& Charv\'{a}t (2009)]{Meteosat}
{Borovi\v{c}ka, J. \& Charv\'{a}t, Z.} 2009.
\textit{A\&A}, 507, 1015

\bibitem[Borovi\v{c}ka \& Spurn\'y (1996)]{Ben96}
{Borovi\v{c}ka, J., \& Spurn\'y, P.} 1996.
\textit{Icarus}, 121, 484

\bibitem[Borovi\v{c}ka et al. (1998)]{Ben98}
{Borovi\v{c}ka, J., Popova, O. P., Nemchinov, I. V., Spurn\'y, P., \& Ceplecha Z.} 1998.
\textit{A\&A}, 334, 713

\bibitem[Borovi\v{c}ka et al. (2013a)]{Kosice}
{Borovi\v{c}ka, J., T\'oth, J., Igaz, A. et al.}  2013a.
\textit{Meteorit. Plan. Sci.}, 48, 1757

\bibitem[Borovi\v{c}ka et al. (2013b)]{Nature1}
{Borovi\v{c}ka, J., Spurn\'y, P., Brown, P. et al.}  2013b.
\textit{Nature}, 503, 235

\bibitem[Borovi\v{c}ka et al. (2015)]{AIV} 
{Borovi\v{c}ka, J., Spurn\'y, P. \& Brown, P.} 2015.
In: P. Michel et al. (eds.), 
 \textit{Asteroids IV} (Tucson: Univ. of Arizona), in press

\bibitem[Brown et al. (2013)]{Nature2}
{Brown, P. G., Assink, J. D., Astiz, L.  et al.} 2013.
\textit{Nature}, 503, 238

\bibitem[Ceplecha \& ReVelle (2005)]{CepReV}
{Ceplecha, Z., \& ReVelle, D. O.} 2005.
\textit{Meteorit. Plan. Sci.}, 40, 35

\bibitem[Ceplecha et al. (1993)]{Cepl93}
{Ceplecha, Z., Spurn\'y, P., Borovi\v{c}ka, J., \& Kecl\'{\i}kov\'a J.} 1993. 
\textit{A\&A}, 507, 1015

\bibitem[Halliday (1968)]{Halliday} 
{Halliday, I.} 1968. In L. Kres\'ak \& P. M. Millman (eds.) \textit{Physics and Dynamics of Meteors} (Dordrecht: D. Reidel),
\textit{IAU Symp.}, 33, 91 

\bibitem[Jenniskens et al. (2009)]{Jenn}
{Jenniskens, P., Shaddad, M. H., Numan, D. et al.} 2009,
\textit{Nature}, 458, 485

\bibitem[Popova et al. 2011]{Popova} 
{Popova, O., Borovi\v{c}ka, J., Hartmann, W. K et al.} 2011.
\textit{Meteorit. Plan. Sci.}, 46, 1525

\bibitem[Popova et al. (2013)]{Science}
{Popova, O. P., Jenniskens, P., Emel`yanenko, V. et al.} 2013.
\textit{Science}, 342, 1069

\bibitem[Pravec \& Harris (2000)]{Pravec}
{Pravec, P. \& Harris, A. W.} 2000,
 \textit{Icarus}, 148, 12

\bibitem[S\'anchez \& Scheeres (2014)]{SS}
{S\'anchez, P.  \& Scheeres, D. J.} 2014,
 \textit{Meteorit. Plan. Sci.}, 49, 788

\bibitem[Scheirich et al. (2010)]{Scheirich}
{Scheirich, P., \v{D}urech, J., Pravec, P. et al.} 2010.
 \textit{Meteorit. Plan. Sci.}, 45, 1804

\bibitem[Shaddad et al. (2010)]{Shaddad}
{Shaddad, M.~H., Jenniskens, P., Numan, D., et al.} 2010.
 \textit{Meteorit. Plan. Sci.}, 45, 1557

\bibitem[Spurn\'y (1994)]{Ben94}
{Spurn\'y, P.} 1994. 
\textit{Plan. Space Sci.}, 42, 157

\bibitem[Spurn\'y et al. (2014)]{Ben14}
{Spurn\'y, P., Haloda, J., Borovi\v{c}ka, J., Shrben\'y, L., \& Halodov\'a, P.} 2014.
\textit{A\&A}, 570, A39

\end{thebibliography}
\end{document}